# RF study and simulations of a C-band Barrel Open Cavity (BOC) pulse compressor

SHU Guan(束冠)[1,2]   ZHAO Feng-Li(赵风利)[1]   HE Xiang(贺祥)[1]

[1]Institute of High Energy Physics, Chinese Academy of Sciences, Beijing 100049, China

[2]University of Chinese Academy of Science, Beijing 100049, China

**Abstract:** This paper focuses on the RF study of a C-band(5712MHz) BOC pulse compressor. The operating principle of BOC is presented and the technical specifications are determined. The main components of BOC such as the cavity, the matching waveguide, the coupling slots and the tuning rings were numerically simulated by 3-D codes software HFSS and CST Microwave Studio(MWS). The "whispering gallery" mode $TM_{6,1,1}$ with an unload Q of 100000 was chosen to oscillate in the cavity. An energy multiplication factor of 1.99 and a peak power gain of 6.34 were achieved theoretically.

## 1  Introduction

The pulse compressor is one of the key devices of linear accelerators. At the moment, only SLED-type and BOC-type pulse compressor exist at C-band[1-4]. The conventional SLED is based upon two high Q resonant cavities which are connected by a 3dB hybrid coupler to minimize the reflection[5]. Contrary to the SLED-type, the BOC-type pulse compressor consists of a single barrel open cavity and a matching waveguide, a special property of the open cavity is its resonant mode—whispering gallery mode, which has an extremely high unload Q, so the BOC-type is more simple, compact and economical compared to the SLED-type.

The study of a BOC was started for the first time in IHEP, a prototype made of OFHC is now under fabrication, which is focused on the research of RF characterizations and mechanic brazing process, and this is an opportunity to gain experience of a rather new RF technology as well.

The performance of a BOC depends on the following three characteristic values: the resonant frequency, the unload Q and the coupling coefficient. HFSS was used to evaluate these parameters and the results are compared with the MWS results as a verification.

## 2  BOC theory and choice of parameters

The BOC is composed of a barrel shape open cavity, a certain number of coupling slots and a waveguide wrapped around the perimeter of the cavity. Fig.1 shows the schematic layout.

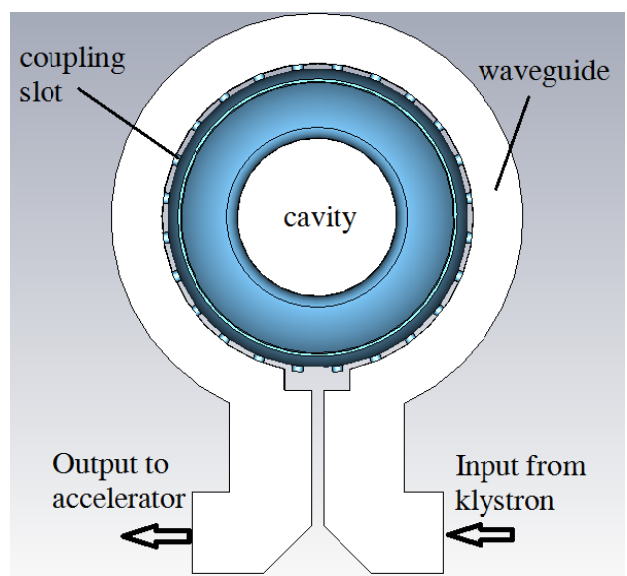

Fig. 1.   The schematic layout of BOC

A fraction of RF power which is magnetically coupled from waveguide to cavity though the dozens of coupling slots propagates around the cavity as a traveling wave. The traveling wave is synchronous with the waveguide. The adjacent coupling slots are spaced $\lambda_g/4$ apart, each slot radiates a forward wave to the accelerator and a backward wave to the klystron. All the forward waves are in phase at BOC output and add up, while the backward waves experience a cancellation, so that only very little power transports back to the klystron. The output of BOC is the superposition of the input wave and the emitted wave which is radiated from the cavity with a 180° phase reversal. BOC stores energy during a large fraction of the pulse length, if the incoming RF pulse phase is reversed 180° at one compressed pulse time, the stored energy is released during a much shorter period, and output peak power is enhanced at the expense of pulse width[6].

Taking into account the performance and fabrication cost of the prototype, $TM_{6,1,1}$ was chosen as the resonant mode with an unload Q factor of 100000. Assuming the input RF pulse length is 3μs and the phase is reversed at 2.67μs, then the relationship between the energy multiplication factor and the coupling coefficient is shown in Fig. 2. The specifications of the accelerator and the input pulse allow a maximum energy multiplication factor of 1.99 when the coupling coefficient β equals to 4.5. The main technical parameters of the designed BOC are listed in Table 1.

Table 1. Technical specifications of the BOC pulse compressor.

| BOC pulse compressor | Design parameter |
| --- | --- |
| Resonant frequency/MHz | 5712 |
| Resonant mode | $TM_{6,1,1}$ |
| Unload Q | ~ 100000 |
| Coupling coefficient(β) | 4.5 |
| RF input pulse length/μs | 3.0 |
| RF compressed pulse length/μs | 0.33 |
| Energy multiplication factor(M) | 1.99 |
| Peak power gain | 6.34 |
| Effective power gain | 4.29 |

## 3 RF design and simulations

### 3.1 Resonant cavity

The characteristics of the open cavity are decided by its internal geometric dimensions. Being reflected by the cavity internal surface repeatedly, the target mode is concentrated in the cavity, while many other parasitic modes radiate out through the two open ports, the spectrum of cavity is quite sparse. The detailed theory of open cavity can be found in some references[7-9]. Fig. 3 shows the electric energy density distribution of the $TM_{6,1,1}$ mode, the power is concentrated near the curved surface.

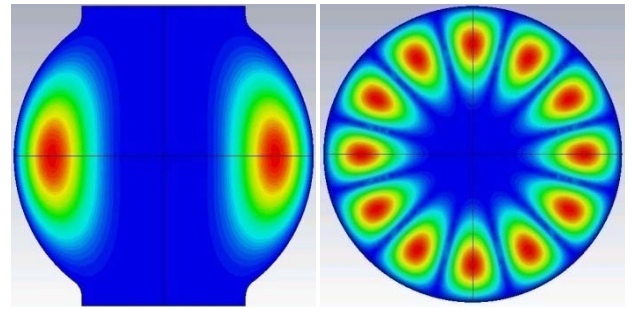

Fig. 3. $TM_{6,1,1}$ mode electric energy density distribution on cut planes simulated by CST.

The "whispering gallery" mode is used in BOC, of which the unload Q value is equivalent to a/δ. The parameter a is the radius of the cavity median plane and δ is the skin depth[7]. In order to obtain a high Q value, the surface flatness should be much smaller than the skin depth.

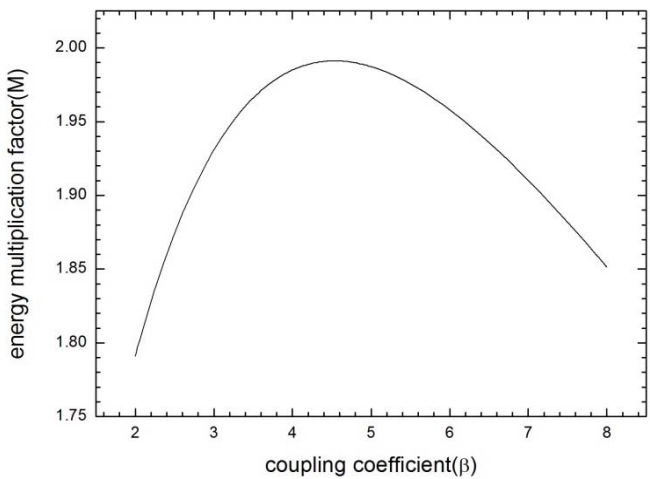

Fig. 2. Relationship between the energy multiplication factor and the coupling coefficient

The cavity was first designed to resonate at a higher frequency (set as 5713MHz) from the target value of 5712MHz to leave some room for frequency adjustment. Table 2 shows the theoretical and calculated value of the BOC cavity. Comparison of the HFSS and CST results with the theoretical ones are with excellent agreement.

Table 2. Theoretical and calculated value of BOC cavity.

| Parameter | Theoretical | HFSS | CST |
| --- | --- | --- | --- |
| Frequency/MHz | 5713.00 | 5713.95 | 5713.94 |
| Unload Q | 100600 | 100300 | 100300 |

### 3.2 Coupling between the cavity and the waveguide

As mentioned in part 2, in order to protect RF source, the distance of adjacent slots is chosen as $\lambda_g/4$ to reduce the backward wave to the klystron. Due to the resonant mode $TM_{6,1,1}$ in the cavity, 24 identical coupling slots are located between the cavity and the waveguide. Phase velocities in the waveguide and cavity are matched by adjusting the width of the waveguide.

The theoretical width of the waveguide is given by[10]:

$$\frac{2\pi(a+\Delta+w)}{6} = \frac{\lambda_0}{\sqrt{1-(\lambda_0/2w)^2}}. \quad (1)$$

where w is the width of waveguide, a is the radius of the cavity median plane, $\Delta$ is the thickness of a slot, $\lambda_0$ is the wavelength of resonant frequency in free space.

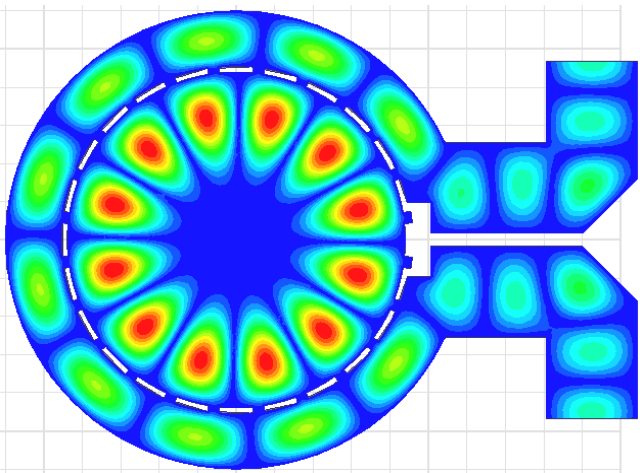

Fig. 4. Electric field distribution on a median plane of BOC pulse compressor simulated by HFSS.

The dimensions of the waveguide were optimized by HFSS and then the whole model simulation was carried out. The electric field distribution on the median plane of BOC pulse compressor is presented in Fig. 4.

The coupling coefficient is determined by the size of slots and the relationship between the coupling coefficient and the radius of the coupling slots is illustrated in Fig.5. For the prototype, the radius is chosen as 3.0mm at first and then enlarged gradually during measurement.

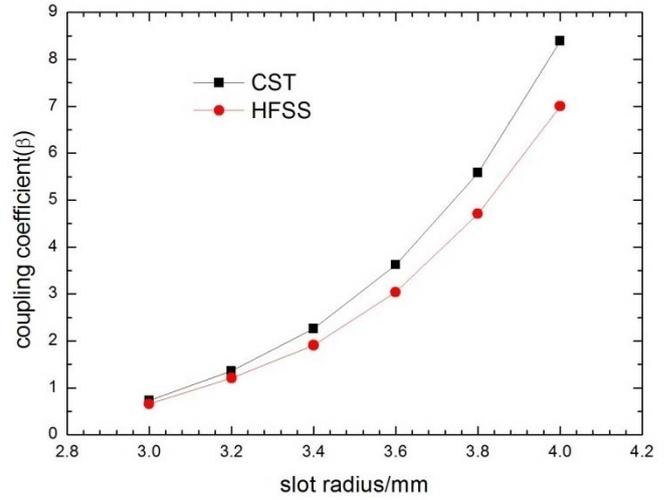

Fig. 5. Dependence of the coupling coefficient on the diameter of the slot

### 3.3 Tuning

Two tuning rings are bulged symmetrical to the median plane in the internal surface of the cavity. The resonant frequency decreases by cutting the rings. Then, the final tuning of the frequency is made later by means of a temperature control using a water cooling system during the BOC operation. Coarse tuning is realized by mechanical cutting, together with temperature control system the frequency is adjusted to 5712MHz eventually[2].

Fig. 6 shows the relationship between the resonant frequency, the unload Q of cavity and the tuning ring radius. The dimensions of the tuning ring were optimized for achieving a large enough tuning range while not affecting the Q-factor of the operating $TM_{6,1,1}$ mode too much. The tuning sensitivity was about 4.4MHz/mm.

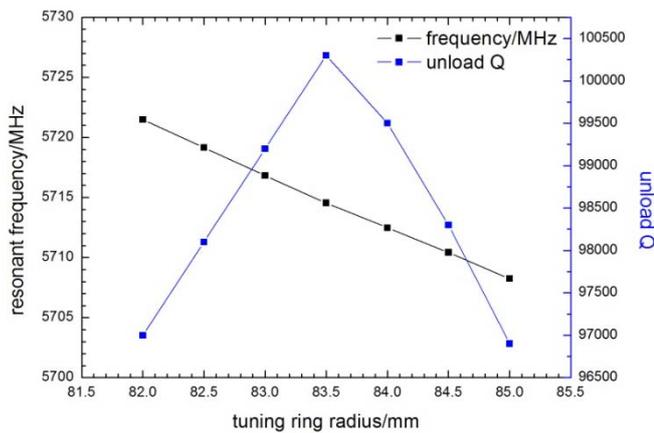

Fig. 6.  Relationship between the frequency, the unload Q and the tuning ring radius simulated by HFSS

## 4  Conclusions

The design of a C-band BOC pulse compressor under the help of HFSS and MWS software is presented in this paper. The BOC prototype is aimed at the research of the RF characterizations in low power, and to check the reliability of the simulations. The study of high power RF electromagnetic effect on the BOC will be carried out later. The cooling channels and the fine tuning systems will be designed according to the RF-thermal-structural-RF coupled analysis.